\documentstyle[prl,aps]{revtex}
\input{psfig}
\input epsf

\newcommand{\ket}[1]{{|#1\rangle}}

\newcommand{\trace}{\mbox{tr}}

\begin{document}

\newcommand{\ooo}{{19}}
\newcommand{\mmm}{{10}}
\newcommand{\xxx}{{121}}
\newcommand{\yyy}{{60}}
\newcommand{\xpyyy}{181}
\newcommand{\www}{{WWW}}
\newcommand{\zzz}{{73205}}
\newcommand{\mzzz}{{163805}}
\newcommand{\uuu}{{98000}}
\newcommand{\vvv}{{220500}}
\newcommand{\ppp}{{190785}}
\newcommand{\qqq}{{415605}}
\newcommand{\offf}{{337195}}
\newcommand{\mfff}{{743215}}
\newcommand{\ottt}{{3.0}}
\newcommand{\mttt}{{1.3}}
%
\newcommand{\orttt}{{313894}}
%
\newcommand{\mrttt}{{838831}}
\newcommand{\oettt}{{146410}}
\newcommand{\mettt}{{327610}}
\newcommand{\pittt}{2.2}
\newcommand{\mpittt}{0.9}

\twocolumn

\title{Resilient Quantum Computation: Error Models and Thresholds}

\author{
Emanuel Knill$^1$\protect\thanks{email: knill@lanl.gov}, Raymond
Laflamme$^2$\protect\thanks{laf@time.lanl.gov}, Wojciech H.
Zurek$^2$\protect\thanks{whz@lanl.gov}}
\address{
 $^1$ CIC-3, MS B265, $^2$ T-6, MS B288
\and
Los Alamos National Laboratory, NM 87545, USA.
}

\maketitle

\begin{abstract}
Recent research has demonstrated that quantum computers can solve
certain types of problems substantially faster than the known
classical algorithms. These problems include factoring integers and
certain physics simulations. Practical quantum computation requires
overcoming the problems of environmental noise and operational errors,
problems which appear to be much more severe than in classical
computation due to the inherent fragility of quantum superpositions
involving many degrees of freedom. Here we show that arbitrarily
accurate quantum computations are possible provided that the error per
operation is below a threshold value.  The result is obtained by
combining quantum error-correction, fault tolerant state recovery,
fault tolerant encoding of operations and concatenation. It holds
under physically realistic assumptions on the errors.
\end{abstract}

The discovery that quantum computers can be much more powerful than
their classical
counterparts~\cite{deutsch:qc1985a,bernstein:qc1993a,simon:qc1994a,shor:qc1994a}
and recent advances in quantum device
technology~\cite{cirac:qc1995a,monroe:qc1995a} have brought the field
of quantum computation into the limelight.  However, until recently,
the hope of taming quantum systems has been overshadowed by the
fragility of quantum information.  Not only must it be preserved in
memory, but it should not be lost when manipulated.  This fragility
comes from two seemingly contradictory requirements.  The system must
be well insulated from the environment to avoid losses, while at the
same time we need to interact with it strongly to perform the desired
computation.
Due to these requirements, it is impossible to completely
isolate a physical quantum computer from the environment.  As a
result, the computer necessarily becomes increasingly entangled with
the outside world and any quantum information it contains is
apparently lost.  Moreover, quantum computation requires the
application of precise unitary operations to the system. It is clear
that these operations cannot be implemented exactly.  For these
reasons, some authors have concluded that the theoretical power of
quantum computers cannot be harnessed~\cite{landauer:qc1995a,unruh:qc1995a}.

The first indication that early assessments of the practicality
of quantum computation might be overly pessimistic was the discovery
of quantum error-correcting codes by Shor~\cite{shor:qc1995b} and
Steane~\cite{steane:qc1995a}. These codes imply that it is possible to
overcome memory or transmission errors provided that the unitary
operators required for encoding, decoding and error-correction could
be implemented with high accuracy. Subsequent work on quantum
error-correction has yielded very efficient codes and generalized much
of the classical
theory~\cite{calderbank:qc1995a,laflamme:qc1996a,bennett:qc1996a,gottesman:qc1996a,calderbank:qc1996a}.
The requirement of accurate error-correction operations was soon
relaxed by Shor~\cite{shor:qc1996a} and essentially eliminated in the
case of quantum channels in~\cite{knill:qc1996b}.  To use quantum
error-correction for quantum computation requires not only encoding
states, but manipulating them in encoded form. That this can be done
in some interesting cases was discovered independently by
Shor~\cite{shor:qc1996a}, Kitaev~\cite{kitaev:qc1996a} and two of us
(W.Z. \& R.L.~\cite{zurek:qc1996a}).  Shor's results in particular
showed that provided the noisy behavior of quantum memory and gates is
stochastic with error probability of the order of
$O(\log^{-\alpha n})$ (where $n$ is the total number of noisy
operations in the quantum network and $\alpha$ is a positive constant),
arbitrarily accurate encoded quantum computation is possible.  
This was a substantial improvement over the best previous bound of $O(1/n)$
and removed almost all fundamental obstacles to practical quantum
computation.

Two obstacles to quantum computation remained. The first is that as
the number of elementary quantum operations grow, Shor's fault
tolerant implementation still require asymptotically zero error per
operation. The second concerns the fact that many if not most error
types expected in real devices cannot be represented in the stochastic
error model. In particular, unitary over-rotation of operations and
small but non-negligible interactions between nearby qubits give rise to
such errors. The purpose of this paper is to show that an error
threshold exists such that if each gate in a physical implementation
of a quantum network has error less than this threshold, it is
possible to perform encoded quantum computations with arbitrary
accuracy. This result holds under physically realistic error models,
thus showing that in principle, unlimited quantum computation with
noise is possible.

In the first part of the paper we introduce the different techniques
required to implement quantum fault tolerance. We emphasize
the various assumptions on errors and then introduce
four fundamental elements of fault tolerance:
quantum error-correcting codes, fault tolerant error-correction
methods, encoded operations and concatenation. We introduce
a new method for obtaining a complete set of fault tolerantly
encodable operations which greatly simplifies the analysis.
We combine the elements to implement a computation fault tolerantly.
In the second part, the fault tolerant networks are analyzed
to obtain rigorous thresholds for the quasi-independent
stochastic and monotonic error models.

\section{Methods}
\label{section:methods}

\subsection{Quantum networks and operations}

\begin{figure}
\begin{center}
\mbox{\psfig{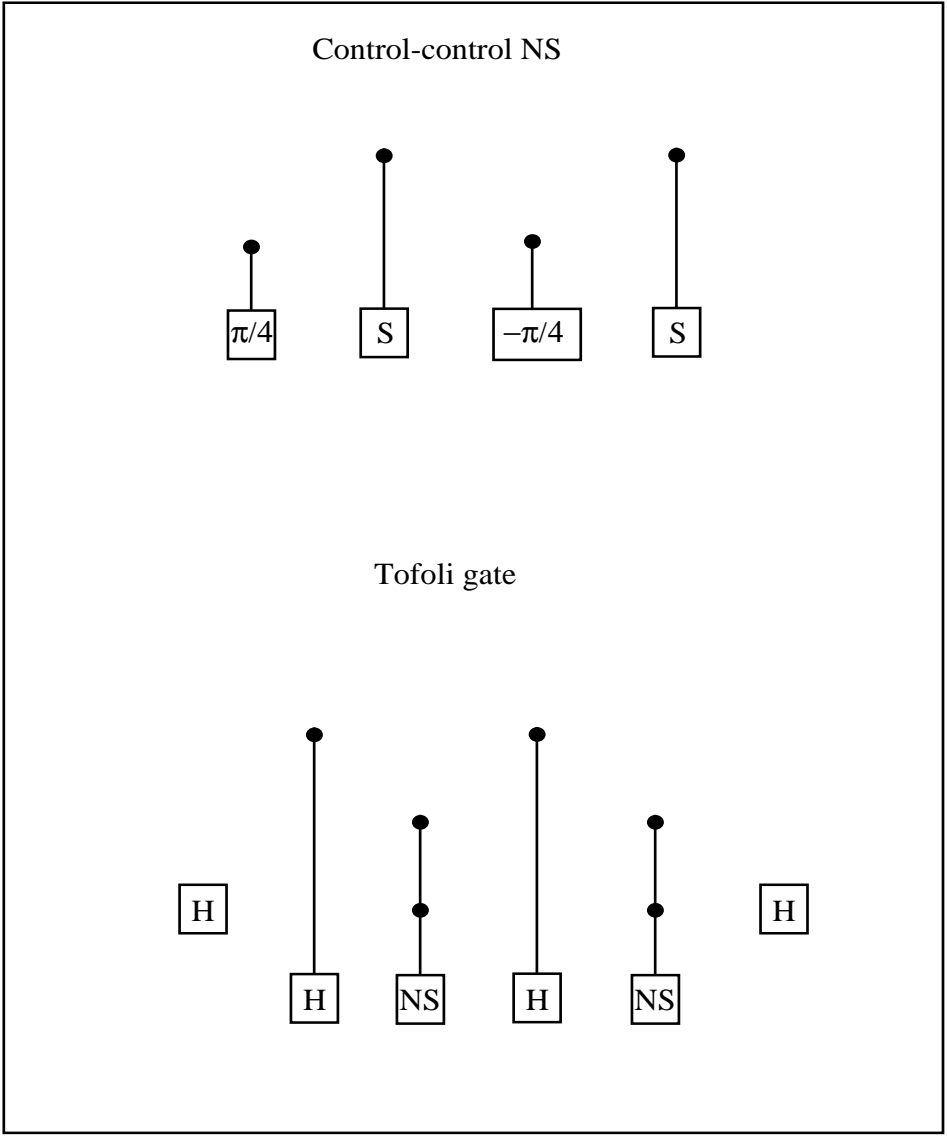}}
\end{center}
\caption{Construction of a Toffoli gate from the primitves given in Table 1.
The first figure gives a controlled-controlled-NS. The second gate is
a Tofoli gate for the first 3 qubit, whatever the 4th bit is.}
\label{figure:toff-by-pi8}
\end{figure}

We can assume without loss of generality that quantum algorithms are
described by means of a quantum network\footnote{ For the purposes of
algorithm design this is not necessarily the best choice. Informal
quantum programming languages (``quantum pseudocode'') are used
in~\cite{cleve:qc1996a,knill:qc1996c}).  }. A quantum network is a
space-time diagram of the operations that are to be applied to each
qubit. Recall that a qubit is the prototypical two state system. It's
state space is the Hilbert space spanned by $\ket{0}$ and $\ket{1}$.
A qubit's time line is represented by a horizontal line, and
operations, that is quantum gates, are denoted by blocks or vertical
lines connected to the qubits being acted on. (See the figures for
examples). A complete set of unitary operations can be formed from
controlled-nots and one qubit phase shifts~\cite{divincenzo:qc1995a}.
A useful set of operations consists of the sign and the bit flip, the
controlled-not, the Hadamard transform and the phase shift of
$\ket{1}$ by $i$. These generate the {\em normalizer group\/}.  To
these one can add the Toffoli gate to achieve
completeness~\cite{shor:qc1996a}. The operations' actions and symbols
are defined in Table~\ref{table:ops}.  To be able to perform quantum
computations requires at least two more operations: preparation of
$\ket{0}$ and measurement of $\ket{0},\ket{1}$ which we call the
{\em classical} basis.  We choose to complete the set of operations by
adding preparation of $\ket{\pi/8} =
\cos({\pi/8})\ket{0} + \sin({\pi/8})\ket{1}$ to our repertoire rather than
the Toffoli gate. A circuit which implements the Toffoli gate using
this operation is given in Figure~\ref{figure:toff-by-pi8}.  Finally,
in many cases, classical computation can be used to process
measurement outcomes and control future quantum gates.  Classical
(reversible, if so desired) circuits may be used to represent these
computations.

The primary purpose of a quantum algorithm is to produce a desired
quantum state from classical input such as
$\ket{0}$. Normally, some or all of the output state's qubits are
measured to obtain the actual information that is needed.  Our main
goal is to describe an implementation of quantum algorithms which
satisfies that if each gate's error is below a threshold value, the
algorithm's computational state is maintained with arbitrary accuracy
in an encoded form. Furthermore, the bits of the computational state
can be measured and the outcome is consistent with the claimed
accuracy of the state.

It is important to realize that the existence and actual value
of the threshold depend critically not only on the details of
the implementation, but also on the assumptions on error
behavior.  The values determined below could be substantially
improved in a particular physcial setting.

\subsection{Assumptions and error models}

To describe a noisy quantum network we introduce the notion of {\em
error locations}. Error locations are to be chosen such that one can
assume without loss of generality that errors ``occur'' only at those
locations, and that these errors satisfy some independence properties
with quantifiable error probabilities or {\em strengths}.

\begin{figure}
\begin{center}
\mbox{\rule{0in}{5in}\hspace{-6in}\psfig{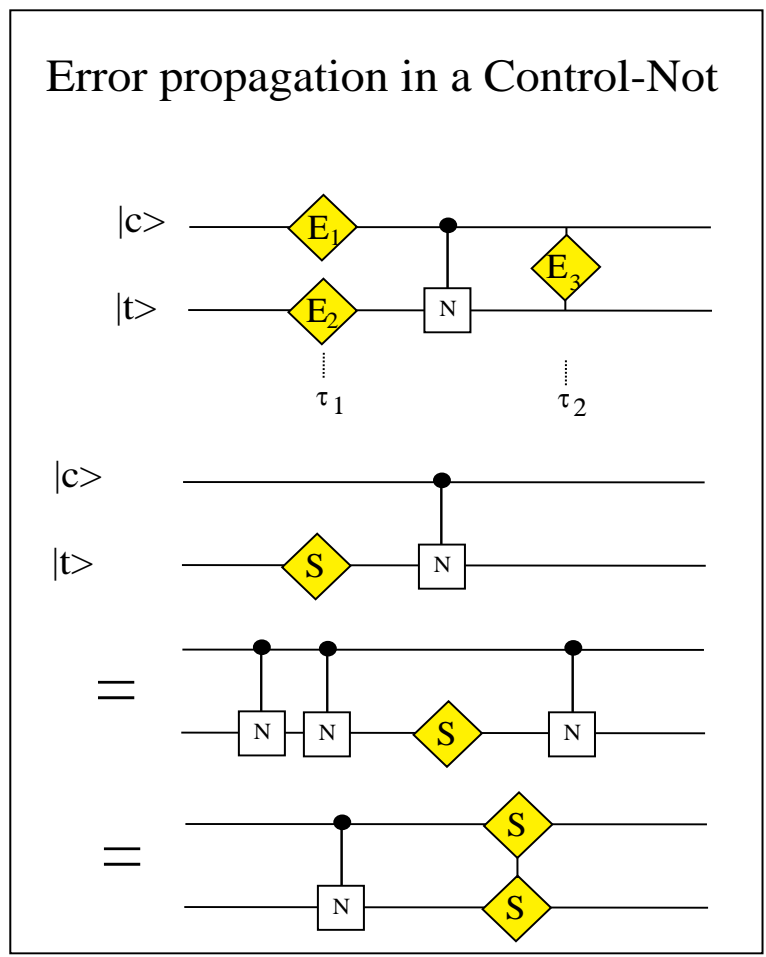}}
\end{center}
\caption{The top part of the figure shows the error location for
a controlled-not gate.  The second part shows the propagation of a sign flip
in the target bit. This type of error propagates from the target to the control
bit.}
\label{figure:error-propagation}
\label{figure:errorlocs-ex}
\end{figure}

We consider two types of error locations: operational error locations
and memory error locations. The former exists after each gate (including
state preparation but not measurements) and
extends over all the qubits involved in the gate's operation.  The
latter exists on each line segment representing a unit time interval
of a qubit's time line, provided the qubit is not involved in a gate
in this interval. The placement of memory errors depends
on the temporal layout of a network and requires partitioning
the network into unit time slices. The time units are determined
by the maximum execution time of a gate. Figure~\ref{figure:errorlocs-ex}
shows how error locations are placed in one of the
component networks required for fault tolerant computing.

We are assuming that sufficiently many gates can be executed
in parallel to avoid loss of quantum information in memory. In fact,
there is a trade-off between memory error and parallelism. If $t$ is
the maximum time a qubit's state can be left in storage without
unacceptable loss of information, then the minimum number of
operations per unit time must be well above $q/t$, where $q$ is the
number of qubits actively involved in computation. If the memory error
per timestep is known and is substantially less than the operational
error, this can be exploited to avoid some parallelism.

The utility of error locations comes from the observation that the
actual behavior of a quantum network can be represented (non-uniquely)
as a mixed sum of networks with linear error operators placed at each
error location. Each network in the sum is associated with a state of
the environment. We call this an {\em error expansion} of the network. The
final state of the computation can be evaluated by obtaining the
states associated with each summand and formally adding them.
It is important to realize that the error expansion
gives correct input-output behavior of the entire network
but need not accurately represent the intermediate physical states of the
qubits.

Assumptions on error behavior are conveniently expressed as
constraints on allowed error expansions.  There are three classes of
constraints that can be considered.  The first involves the types of
error operators that can occur at a given location.  We will for the
most part assume that these operators do not lead to loss of amplitude
from the two dimensional computational subspace in a qubit ({\em
leakage errors}).  We will briefly discuss how this assumption can be
relaxed by exploiting special gates which reliably return lost
amplitudes to the computational subspace.  With the assumption
that there are no leakage errors, we
can, without further loss of generality, take the error operators to be
one of the standard errors (no error, bit flip, sign flip or both).
An additional assumption we will use is that errors in classical
computations based on measurement outcomes are error free. This is a
good approximation to what is possible in practice. We will return to
this assumption in the concluding comments.

The second class of constraints involves the nature of the
mixture. The analysis of our methods is simplest when the mixture is
obtained by stochastically and independently placing one of the
standard errors at each error location. Given our assumption on the
error operators, this is equivalent to requiring that the states of
the environment associated with each summand of the error expansion
are orthogonal. A weaker assumption requires only that a stochastic
error expansion exists without making restrictions on the operators
involved. In this case stochasticity is defined
as having orthogonal environments associated with each
summand. For such expansions it makes sense to define the probability
of error at a given location as the probability of those summands in
the expansion which do not have an identity operator at this location.
Note that these probabilities are in principle state dependent.
Since the input states are in effect predetermined in
a quantum network, the probabilities are fixed but may be
difficult to obtain by calculation.

The third class of constraints enforces independence of
errors at different locations. With this in mind
we define the following error models:
\begin{itemize}
\item[$\bullet$]Independent stochastic errors:
The error expansion is obtained by making an independent, random
choice at each error location of either the identity operator or a
quantum operation. A quantum operation is defined 
as a stochastic (in the sense defined above)
combination of linear operators satisfying a unitarity
condition~\cite{knill:qc1995e,schumacher:qc1996a}. The probability of
error at a given location is defined as the probability of making the
latter choice. The maximum probability of failure over all error
locations is the error probability associated with the model. A convenient
strengthening of this error model requires that the quantum operation
consists of a random choice of one of the standard errors.
\item[$\bullet$]Independent errors:
The error expansion is obtained by assigning a quantum
operation to each error location. In this case we can
no longer talk of error probabilities. The error {\em strength}
of a quantum operation is defined below. The error strength
associated with this model is the maximum error strength
of the quantum operations.
\item[$\bullet$]Quasi-independent stochastic errors:
An error expansion satisfies the quasi-independent stochastic
error model with error probability $p$ if it can be decomposed
into a stochastic sum with the property that the probability
of all summands which have a non-identity quantum operation
at a given $k$ many error locations is at most $p^k$.
\item[$\bullet$]Quasi-independent errors:
Let each summand of an error expansion be associated with a set of
{\em failed} error locations such that all other error locations are
instantiated with an identity operator.  Such an error expansion
satisfies the quasi-independent error model with error strength $p$ if
the total strength of the summands for which at least a given $k$ many
error locations have failed is at most $p^k$.
It satisfies the quasi-independent monotonic error model
with bound $C$ if the total strength of any subset
of these summands is at most $C p^k$.
\end{itemize}
The analysis to be presented below establishes thresholds
for quasi-independent stochastic  and monotonic
errors. The analysis for quasi-independent
errors without the monotonicity assumption is more complicated
and yields a somewhat worse threshold. It will be presented
elsewhere.

The strength of a set of linear operators labeled by 
states of the environment is defined as follows:
\[
|\sum_i\ket{e_i}A_i| = \sup_{\ket{\psi}}|\sum_i\ket{e_i}A_i\ket{\psi}|.
\]
This is the maximum modulus of the amplitude of the outcome of
applying the operation to a (normalized) state.  Given a
representation of a quantum operation of the form
$\ket{e_0}I+\sum_i\ket{e_i}A_i$, its error strength is the strength of
the second term.  Any quantum operation has a representation of this
form with the property that $\trace A_i = 0$ for each $i\geq 1$.  By
default, the error strength is determined by such a representation In
this case, the error strength is independent of the choice of
representation with this property.  Note that in the case of summands
of error expansion, we could, at least in principle, replace the
supremum in the definition of error strength by evaluation at the
known input state. Error strength is closely related to the standard
notion of fidelity~\cite{knill:qc1995e,schumacher:qc1996a} but easier
to analyze in the present context.  For stochastic models, error
probability is given by the square of the error strength. Thus noise
limits are generally much more stringent when considering
non-stochastic models.

The error models described above can be used to bound the
probability\footnote{
When discussing both stochastic and non-stochastic models,
we use the word ``probability'' to mean either probability
or strength, depending on the model.}
of a network's computation failing. This is at
most the probability of at least one error location having
a non-identity operator in the error expansion. Suppose the network
has $n$ error locations. Then for the quasi-independent stochastic or
monotone models with error probability $p$, this is at
most $np$ or $Cnp$, respectively. For the quasi-independent error model
with error strength $p$ this is given by $(1+p)^n - 1$~\footnote{
This bound is obtained by inclusion-exclusion based on the
known bounds for summands involving failures at given locations.}.
Thus high probability of success is assured if $p\ll 1/n$.

How realistic are these error models? Since physics is generally
described using local interactions, it seems reasonable to assume that
error events on qubits are caused by independent environments except
when they are intentionally modified by an interaction implementing a
multi-qubit operation.  This is the independent error model in
physical terms. It covers not only independent stochastic errors, but
also non-identity unitary operators at each error site.  For example,
a consistent modification of the internal energies of each qubit by a
weak external field is allowed, provided the deviations are small
enough. The quasi-independent error model is substantially more
general than the independent error model. In addition to other error
types, it can also model weak pairwise interactions between adjacent
qubits. The quasi-independent monotonic error model makes the
reasonable additional assumption that relevant sums of subsets of
operators in an error expansion do not conspire to greatly increase
the modulus of the amplitude of the input state.

The assumption that no amplitude is lost from the computational
systems can be quite unrealistic. For example in the
ion trap, amplitude may be lost to any of the other available
levels, some of which are in fact needed for some operations
and readout. Amplitude loss can be modeled with error expansions
by extending the dimension of each system and allowing
non-computational error operators. However, the fault tolerant
networks implemented here can fail in the presence of these errors.
To make them work requires the use of {\em stop leak gates\/}
as described in the analysis section.

Although the above arguments suggest that the quasi-independence
assumptions can be satisfied in principle, it is worth
pointing out that on the surface, some proposed implementations such
as the ion trap quantum computer, violate this assumption from the
beginning by involving a shared bus in every two-qubit operation.
To ensure that independence can still be assumed requires
reliable dissipation of residual information in the bus
qubit between any two operations using it. The need
for such dissipation is similar to the need for reliable
methods to restore amplitude to the computational state
space of a qubit in the case of leakage errors.
Dissipation of bus qubit information is used in~\cite{cirac:qc1996a}
for controlling bus errors.

\subsection{Quantum error-correction}

The first element required for fault tolerance is quantum
error-correction. In classical communication and computation,
error-correction is usually accomplished by redundantly encoding
information. The simplest and quite effective if inefficient method
involves duplicating the information at least three times and using
majority voting to recover it after errors have occurred.  This method
cannot be straightforwardly applied to quantum computers for three
reasons.  First it is not possible to clone arbitrary quantum
states\cite{wooters:qc1982a}.  Second, in order to take a majority
vote, it is naively thought that we must first learn the encoded
information by measurement.  This would destroy any quantum coherence
of the state. Finally, only one type
of error needs to be considered for classical binary information,
the bit flip. Quantum states can be modified by a continuum
of possible errors.

Recently Shor\cite{shor:qc1995b} and Steane\cite{steane:qc1995a,steane:qc1995b}
discovered how these objections could be overcome.  To avoid copying
information to introduce redundancy it is possible to exploit highly
entangled states supported by additional qubits.  Thus a quantum state
is unitarily associated with a linear combination of such entangled
states on sufficiently many qubits to permit recovering the
information after the loss of information in few of the qubits. These
linear combinations define the coding subspace.  To avoid collapse of
the quantum information by measurement in the process of correcting
errors, it is possible to make a partial measurement which extracts
only error information and leaves the encoded state untouched.  This
can be arranged by choosing an encoding with the property that the
standard errors translate the coding subspace to orthogonal subspaces.
To deal with the fact that there is a continuum of possible errors, it
is sufficient to recognize that every error can be represented as a
linear combination of the standard errors. Together with the
observation that linear combinations of correctable errors are also
correctable, this allows discretizing the error possibilities.  The
general theory of error-correction is discussed
in~\cite{knill:qc1995e}. Standard errors are introduced
in~\cite{steane:qc1995a}.

To prove the
threshold theorem we make use of the one qubit single error-correcting
code on seven qubits based on the classical Hamming code
\cite{steane:qc1995a,calderbank:qc1995a}.  One can
view this code as a prescription for interpreting the seven qubit
system as a pair of abstract particles, the abstract qubit and the
syndrome space. Thus $Q\otimes Q\otimes Q\otimes Q\otimes Q\otimes
Q\otimes Q \simeq A\otimes S$, where $A$ is the abstract two state
particle and $S$ is the syndrome space. The syndrome space is endowed
with a basis $\ket{000000_S},\ket{000001_S},\ldots,\ket{111111_S}$
($64$ orthonormal states). The one qubit state
$\alpha\ket{0}+\beta\ket{1}$ is encoded as
$\ket{\psi}=(\alpha\ket{0_A}+\beta\ket{1_A})\ket{000000_S}$ in $
A\otimes S$. The representation ensures that if an error operator acts
on one of the seven supporting qubits, the state component of
$\ket{\psi}$ in $A$ is unchanged and only the syndrome space is
affected. This permits recovery of the state $\ket{\psi}$ by measuring
$S$ and resetting its state to $\ket{00000_S}$. Provided that this can
be done sufficiently frequently and reliably, one can maintain the
state of the abstract particle $A$ for prolonged periods, even in the
presence of noise acting independently on the seven qubits of the
code.

The representation $A\otimes S$ can be specified by
supplying observables for $A$ and $S$. We provide
six binary observables for $S$ 
with eigenvalues $\pm 1$ depending on the value
of the corresponding bit in the labeling of the basis.
These observables  are given by
\begin{eqnarray*}
S_1 &=&
I\otimes I\otimes I\otimes\sigma_z\otimes\sigma_z\otimes\sigma_z\otimes\sigma_z,
\\
S_2 &=&
I\otimes \sigma_z\otimes\sigma_z\otimes I\otimes I\otimes \sigma_z\otimes\sigma_z,
\\
S_3 &=&
\sigma_z\otimes I\otimes\sigma_z\otimes I\otimes\sigma_z\otimes I\otimes\sigma_z,\\
S_4 &=&
I\otimes I\otimes I\otimes\sigma_y\otimes\sigma_y\otimes\sigma_y\otimes\sigma_y,
\\
S_5 &=&
I\otimes \sigma_y\otimes\sigma_y\otimes I\otimes I\otimes \sigma_y\otimes\sigma_y,
\\
S_6 &=&
\sigma_y\otimes I\otimes\sigma_y\otimes I\otimes\sigma_y\otimes I\otimes\sigma_y.
\end{eqnarray*}
An observable for $A$ which behaves correctly relative to the errors we
wish to protect against is not as easy to provide, due to its
dependence on a non-linear syndrome decoding method.  Instead we
provide the following observable which defines the correct encoding
subspace, and for given values of the syndrome observable gives a
basis related to the abstract particle's natural basis by one of the
standard errors:
\[
A_1 =
\sigma_z\otimes\sigma_z\otimes\sigma_z\otimes\sigma_z\otimes\sigma_z\otimes\sigma_z\otimes\sigma_z.\]
Which standard error affects this observable given
the state of the syndrome space can be determined
from the details of the decoding method.

The fact that this code corrects one qubit errors is established
in~\cite{steane:qc1995a}.  Each of the syndrome's observables can be
measured by a simple circuit which determines a parity in either the
$\ket{0},\ket{1}$ or the $\ket{+}\ket{-}$ basis of the qubits. 
The parity checks yield the
syndrome state. Each combination of standard errors applied to the
qubit changes the values of the syndrome observables in a
deterministic fashion.  Each error consisting of at most one bit flip
and one sign flip yields a distinct syndrome state. Thus, which of these
errors occurred can be learned from the syndrome, and the error can be
corrected (up to a global phase) by applying the same operators again.

\subsection{Fault tolerant error-correction}

\begin{figure}
\begin{center}
\mbox{\rule{0in}{7in}\hspace{.3in}\psfig{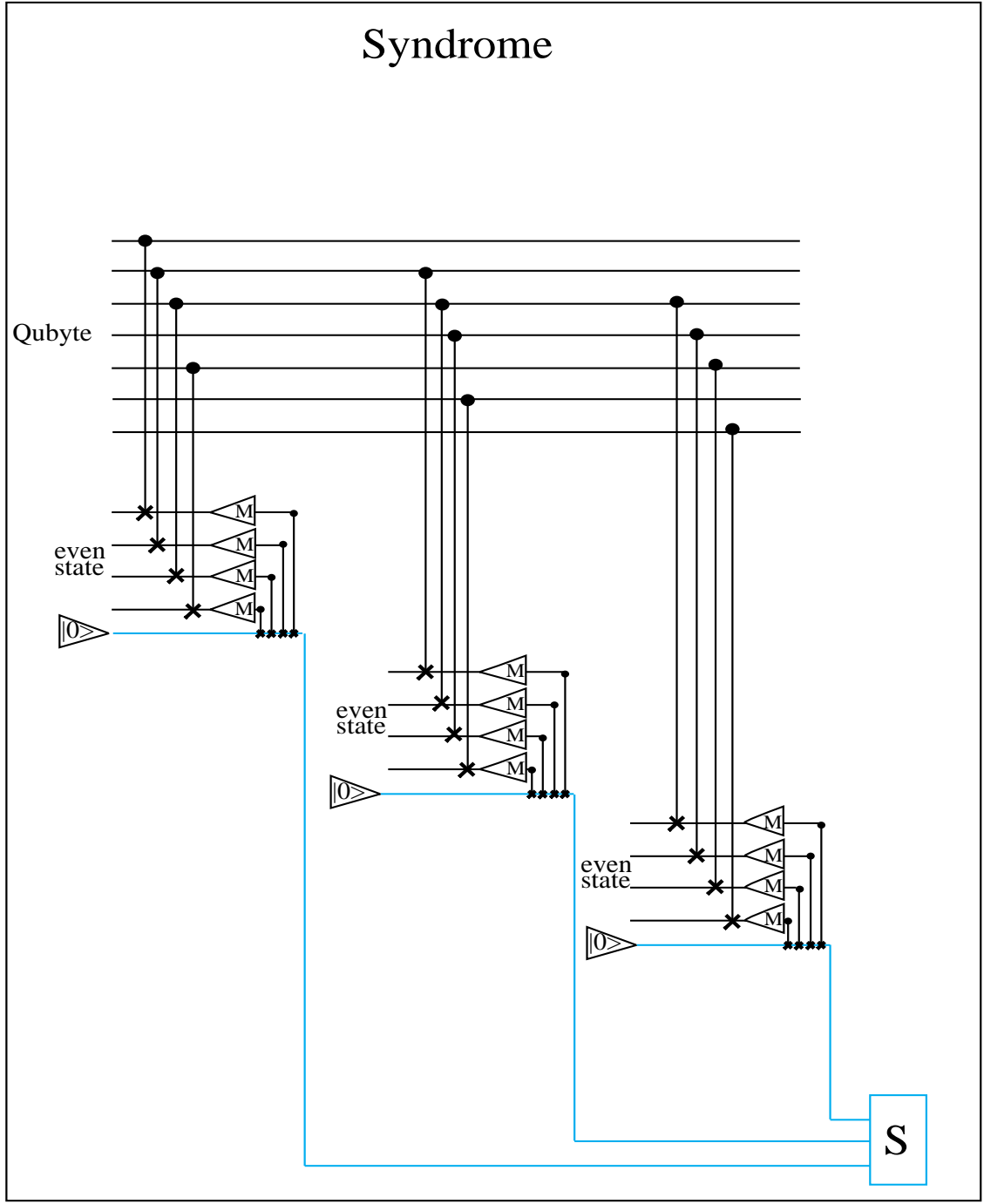}}
\end{center}
\caption{Fault tolerant calculation of the syndrome 
for the 7 bit code.  The even state is obtained from the fault tolerant
operation given by [shor]}
\label{figure:parity_checks}
\end{figure}

\begin{figure}
\begin{center}
\mbox{\rule{0in}{4in}\hspace{-1in}\psfig{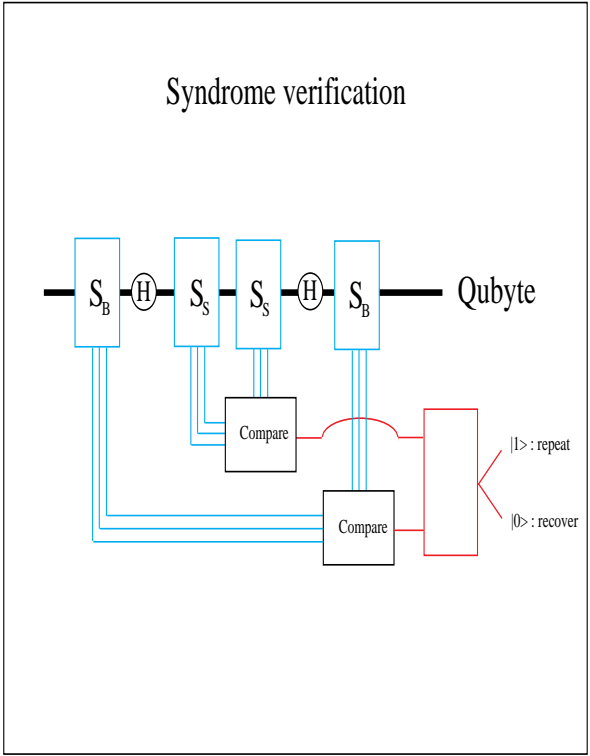}}
\end{center}
\caption{Verification of the syndrome of a qubyte. The syndrome
for both  bit flips $S_B$ and sign flips  $S_S$
are calculated twice to learn if the error occured during the syndrome oepration.}
\label{figure:ft-recovery}
\end{figure}

Error-correcting codes work under the assumptions that the encoding
and recovery (the step consisting of detecting the errors and
correcting them) are error free.  
This can be a good approximation in the case of quantum
communication or memory where the errors occur mostly during
transmission or idle time and not in the encoding and recovery period.
In general, we cannot expect that the operations involved in
recovering the encoded state are exact and a method must be devised
for implementing them such that any single error (in the case of the
7-qubit code) occurring anywhere in the network can be eliminated. An
efficient method for accomplishing this goal is given
in~\cite{shor:qc1996a,divincenzo:qc1996a}. It is based on two basic
ideas. The first is to use prepared, verified ``cat'' states to permit
extracting the syndrome information without directly applying
multi-qubit operations to the qubits of the code. The verification of
the prepared states is used to reliably eliminate pre-existing errors
which could adversely affect more than one qubit of the code when the
syndrome is extracted. The second idea is to perform the syndrome
extraction multiple times to ensure that the syndrome has been
measured correctly. This avoids the problem of destroying the state by
failing to restore the syndrome to $\ket{000000_S}$. Note that the
operators involved in restoring the syndrome act independently on each
qubit. A network for fault tolerant recovery suitable for analysis is
shown in Figures~\ref{figure:parity_checks}
and~\ref{figure:ft-recovery}.  To explicitly check that the method
described has the desired effect of tolerating at least one error in
the error locations of the network requires a detailed study of how
errors propagate. Error propagation is discussed in
Section~\ref{section:analysis}.
\smallskip

\subsection{Encoded operations}

Quantum error correcting and fault tolerant methods 
protect quantum information when the computer is not actively
performing logical operations on the computational state. A quantum
computation will be required to perform such operations.
This can of course be done by
decoding the state, applying the desired operation and reencoding
it. However, this temporarily removes the protection offered by the
encoding, a serious problem if one wishes to compute fault tolerantly.
Thus it is necessary to act directly on the encoded state by means of
suitably encoded operations. The trick is to do this without having
any single operational error introducing multi-qubit errors in (for
example) the seven qubits used for encoding an abstract particle.  The
simplest method for accomplishing this involves implementing
operations {\em transversally}.  Thus we label the qubits of each
coded qubit from $1$ to $7$ and permit interactions between them only
if they have the same label. Note that with a suitable choice of
labels on the even states, this rule is satisfied by the recovery
operator, but not by the even state preparation procedure. In the
latter case, dangerous two or more qubit errors are eliminated by the
verification step.

\begin{figure}
\begin{center}
\mbox{\psfig{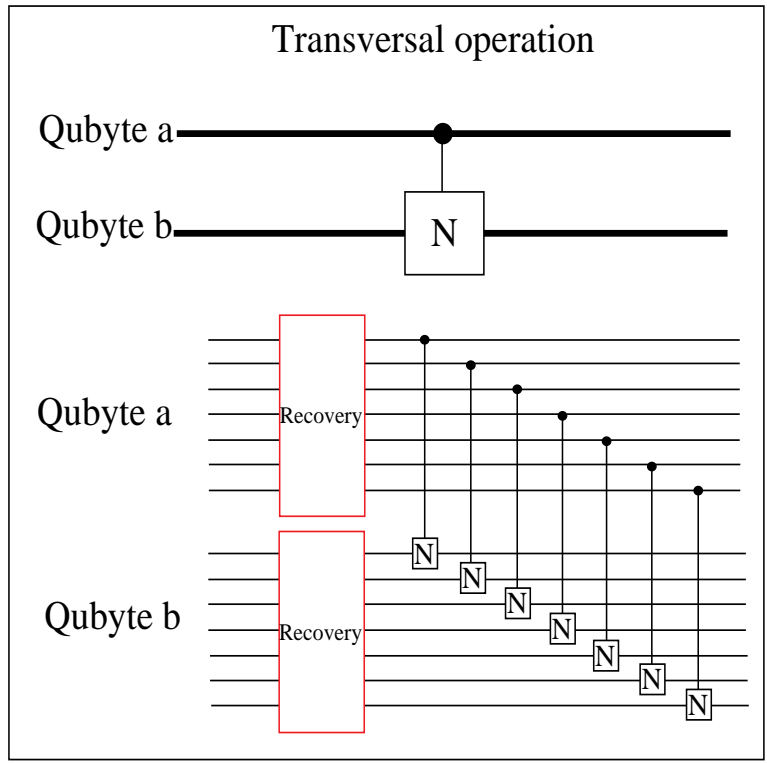}}
\end{center}
\caption{Fault tolerant implementation of a control not between two qubytes.
Each qubit of a qubyte interacts at most with one qubit of another qubyte:
this is the transversality property.}
\label{figure:normalizer_implementation}
\end{figure}

What operations on states encoded using the seven qubit code
can be performed transversally?
In~\cite{shor:qc1996a} it is shown that the normalizer group as well
as preparation of encoded $\ket{0}$ and classical measurement can be
implemented in this fashion. 
Such a network for the controlled-not
is shown in Figure~\ref{figure:normalizer_implementation}.
Similar implementations are possible for many other codes
as demonstrated by Gottesman~\cite{gottesman:qc1997a}.

A useful property is
that encoding, decoding and recovery operators can be implemented
using only generators of the normalizer group. Measurement is
performed by measuring each qubit in the classical basis and decoding
the outcome classically. Preparation involves fault tolerantly
measuring the observable $A_1$ in addition to the syndrome and
resetting all observables to $0$ by applying a suitable combination of
standard errors. 

\begin{figure}
\begin{center}
\mbox{\rule{0in}{4in}\hspace{-8in}
\psfig{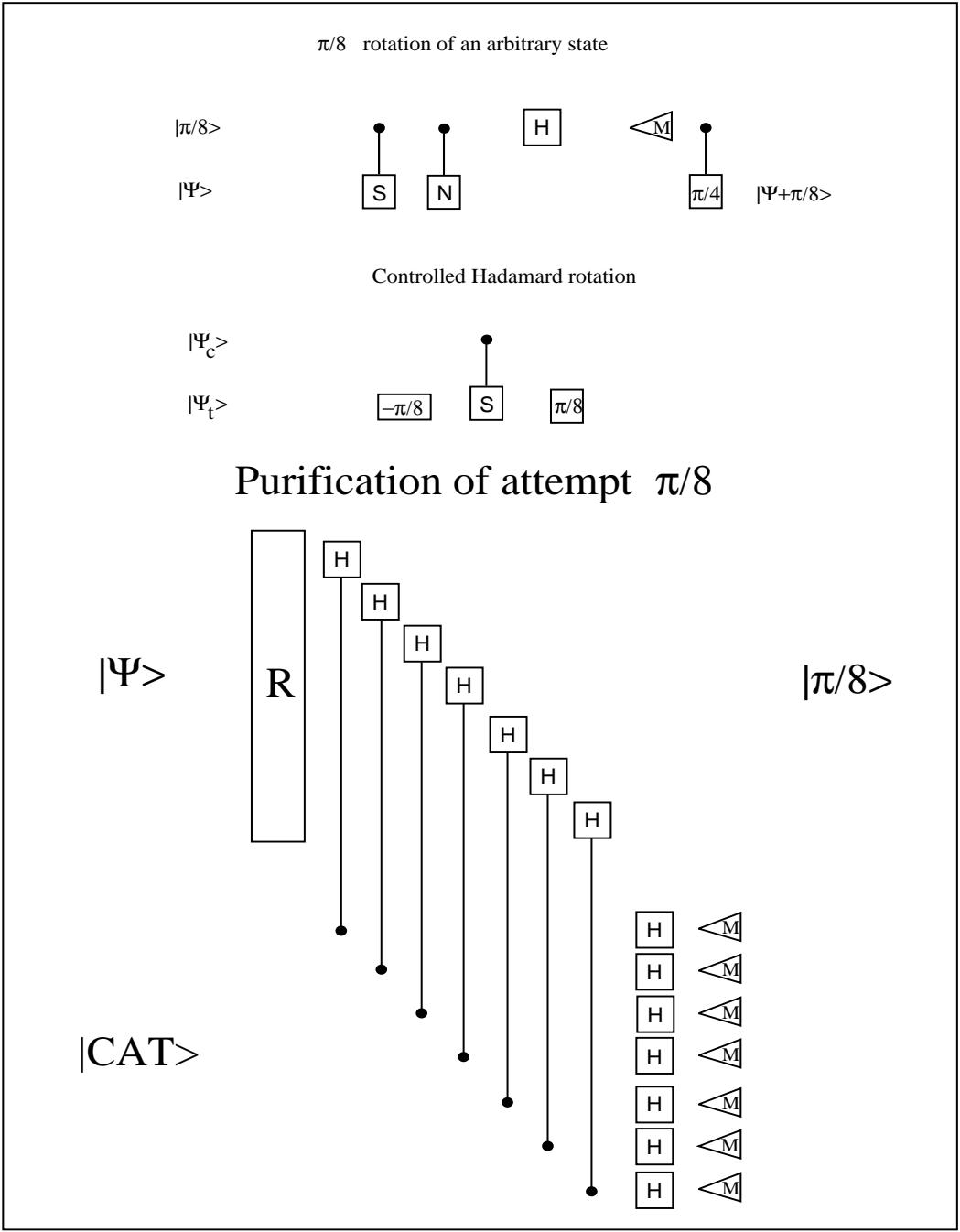}}
\end{center}
\caption{Fault tolerant network giving purified $|\pi/8\rangle$ states.}
\label{figure:ft-pi8}
\end{figure}

The normalizer group is not computationally
complete. In~\cite{shor:qc1996a}, a fault tolerant method for
implementing the Toffoli gate is used to achieve completeness and
in~\cite{knill:qc1996j} two conceptually simpler techniques involving
two-qubit operations are given. Here we use a third idea which requires
only preparation and verification of the $\ket{\pi/8} =
\cos(\pi/8)\ket{0}+\sin(\pi/8)\ket{1}$
state. Figure~\ref{figure:ft-pi8} shows networks for generating a
reliable encoded $\ket{\pi/8}$ state given the ability to prepare
unencoded $\ket{\pi/8}$ states.  The trick is to realize that
$\ket{\pi/8}$ is the $+1$ eigenstate of the Hadamard transform. By
performing a measurement using Kitaev's
techniques~\cite{kitaev:qc1996a} via a controlled Hadamard transform,
one can obtain either $\ket{\pi/8}$ or $\ket{5\pi/8}$. The latter can
be converted into the former by a $\pi/2$ rotation.
The controlled Hadamard transform can be implemented transversally
as shown in Figure~\ref{figure:ft-pi8}. To eliminate single
errors, the measurement is performed twice with an intermediate
recovery network. The state is rejected if the two measurements
do not agree.

State preparation and measurement are important not only for
generating and exploiting $\ket{\pi/8}$, but for computation in
general. As shown in~\cite{aharanov:qc1996c}, without the ability to
generate bounded noise initial states at any time during the
computation, efficient fault tolerance is unachievable. Intuitively,
this can be explained by viewing fault tolerant recovery as a method
for maintaining the error ``temperature'' of the computational state
by transferring error information to the cooler states such as the
prepared even states.  Measurement is useful for  accessing the
error information and removing its entropy and for input/output.

\subsection{Concatenation}

\begin{figure}
\begin{center}
\mbox{\psfig{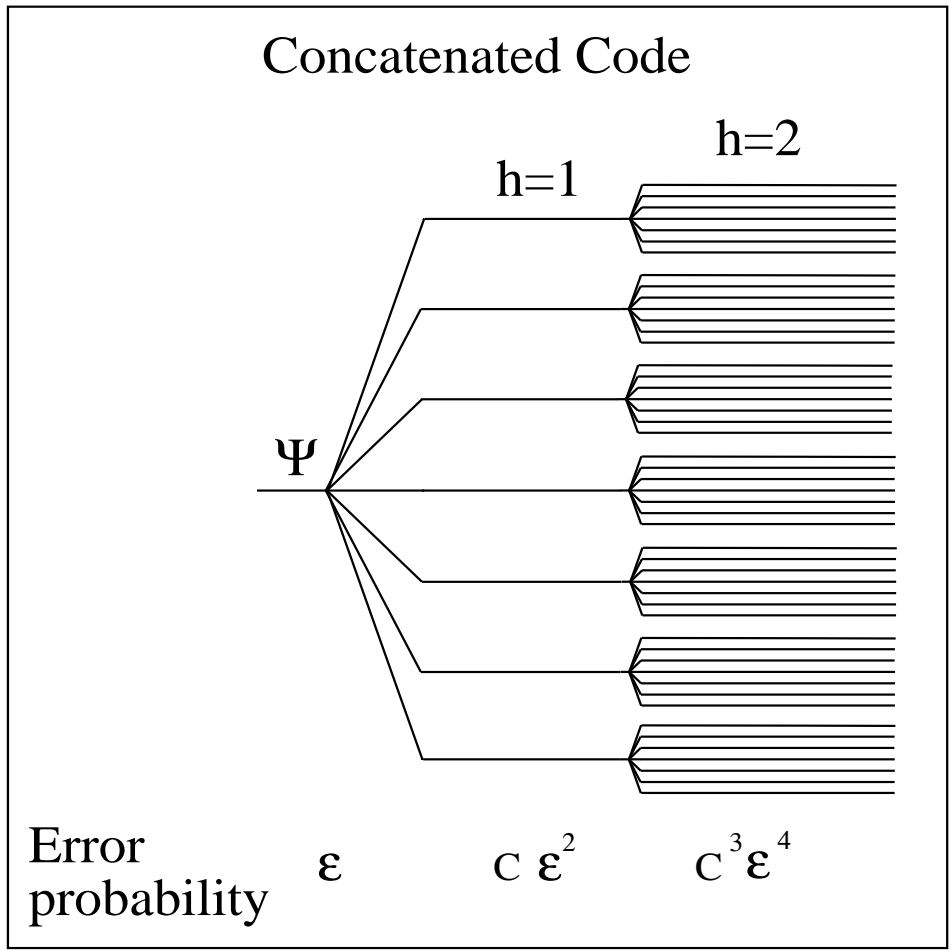}}
\end{center}
\caption{Concatenation of the 7 bit code. If the error rate is $\epsilon$
for the qubits, the encoding will gives a rate of $C^{2^h-1}\epsilon^{2^h}$ for the $h^{th}$ level of the hierarchy.}
\label{figure:concatenation}
\end{figure}

The final ingredient required to implement fault tolerance with a
threshold error rate is concatenation. The combination of quantum
error-correction, fault tolerant error-correction and encoded
operations can be viewed as a technique for exploiting qubits on which
we can operate with an error $p$ per operation to define abstract
(encoded) qubits with a smaller error per operation. With the
implementations described above, the effective error probability or
strength is reduced (for sufficiently small $p$)
from $p$ to $cp^2$, where $c$ is to be determined
below.  Concatenation involves applying the same combination of
techniques using the abstract qubits as the starting two state systems
for encoding.  Figure~\ref{figure:concatenation} shows the networks
obtained by applying concatenation twice. We shall see that the
effective error on the encoded state after $h$ iterations of the
concatenation procedure is reduced to $c^{2^h-1}p^{2^h}$.

\section{Analysis}
\label{section:analysis}

There is a simple method for estimating accuracy thresholds associated
with concatenation. In many cases, the threshold is obtained by
exploiting ``single-error elimination'' networks, which have the
property that if an error occurs at an error location with no other
errors present in some region of influence, then this error will not
affect the computational state.  A naive estimate for the probability
of error at the next level is given by the number of pairs of errors
that can occur within the region of influence times the square of the
probability of error at this level.  In networks such as those used in
this work, the number of error locations in the region of influence is
bounded by the number of qubits required for the largest gates (here
this is the controlled-not, with 14 qubits), times the number of
operations affecting a qubit before an error at a given location can
be eliminated (here this is estimated as $7*6*2+2$, where $7$ is
(close to) the average number of operations contributing to a qubit in
the cat state generator, $6$ is the number of syndrome bits, $2$ is
the average number of attempts to measure the syndrome, and the
additional $2$ is for the correction step and the encoded operation).
For our network we get $14*86=1204$ operations, which yields
a probability of error at the next level of less than $10^6 p^2$, which
is less than $p$ if $p<10^{-6}$.  This is close to what will be
established by the more formal arguments below.

Our analysis consists of establishing the error behavior 
inductively for each level of the concatenation hierarchy.
To do so we exploit the properties of stabilizer codes
and the quasi-independent error models.
We first prove the threshold theorem for the
normalizer group and the quasi-independent stochastic
error model. We then
show how it can be extended the monotone model
and to a complete set of operations.

Useful properties of stabilizer codes are:
\begin{itemize}
\item[(i)] Recovery operations and
encoded normalizer operations are implemented using normalizer
operations only.
\item[(ii)] The syndrome is completely
determined by the standard error operations applied to
the qubits.
\item[(iii)] If the errors in different error locations
during fault tolerant recovery are instantiated by standard errors and
the incoming state's syndrome is given, then
the outgoing state's syndrome is determined.
\item[(iv)] Standard errors applied to an encoded state
whose syndrome is given induce a standard error on the encoded state.
\end{itemize}

\subsection{Error propagation}

Recall that an instance of the error behavior of a computation is
obtained by specifying one of the standard errors at each error
location. The analysis of errors at the next level requires
determining the effect of these standard errors on the state of the
syndrome between gates. This is accomplished by moving errors from
locations internal to the encoded gates to locations between gates by error
propagation.  The basic technique is to conjugate standard errors
through gates as illustrated in Figure~\ref{figure:error-propagation}.
A complete list of conjugation relationships for standard errors and
normalizer operations is given in
Table~\ref{table:conjugation-errors}.

It is not possible to propagate errors through measurement
operations. Note however that sign flip errors do not affect the
measurement outcome, while bit flip errors flip the classical outcome.
Given the syndrome before a recovery operation and the standard errors
at error locations in the recovery operation, the classical parities
that are obtained by the measurements are determined and the effect of
the recovery on the state can be computed.  In this case all errors in
the recovery network can be replaced by errors before and after the
recovery network.

Given the standard errors at all error locations, the syndromes of the
encoded states between gates can be determined by first computing them
for the encoded state preparations and then applying the above property
of the recovery network to each gate whenever the input syndromes have
been obtained.  In principle this induces an error behavior at the
next level which can then be analyzed without further reference to the
syndromes at this level.  Unfortunately, this simple approach does not
easily permit computing the relevant error strengths at the next
level.  To do that we use additional error propagation steps to assure
quasi-independent behavior.

\subsection{Encoded gate failures}

For the purposes of analysis, we refine our error models so that
each summand of the error expansion is also associated with
a labeling of each gate or memory operation which declares whether or not
that operation {\em failed}. The labeling is assumed to satisfy
that any operation whose associated error-location has a non-identity
operator has failed, but the converse need not hold.
The induction hypothesis requires that the error model
(quasi-independent stochastic or monotone) applies to the
summands associated with failures at any given $k$ error locations.
Each of the error models defined earlier permits a failure
labeling such that this holds. The goal of the analysis
is to show that the hypothesis holds at the next level
for a quadratically smaller error parameter.

Recall that an encoded gate consists of a (extended, for preparation
gates) recovery network for each input encoded state and a
transversally encoded operation.  Our algorithm for determining
whether an encoded gate failed and establishing the induced behavior
with errors depends only on the failure locations at the current
level, not on the actual standard error associated with the location.
This implies that the analysis is correct provided that the failure
events behave stochastically. Different standard errors associated
with the same failure patterns can combine non-stochastically.
In principle this permits a similar analysis to be used for
situations where the standard errors do not commute with the network
operations.

The first step is to determine for each recovery used in a state
preparation or preceding a gate, whether it was successful. It is
considered successful if after a suitable error propagation step, the
correct syndrome has been obtained.  Recall that
each recovery network consists of two recovery attempts, where the
second one is executed only if the first one has failed due to an
inconsistency in the measured parities.  Each attempt involves two
extractions of the syndrome using prepared cat states. The following
can be established by a careful analysis of the recovery network.
\begin{itemize}
\item[(i)] If no failure occurs in the first recovery attempt,
then the correct
syndrome has been obtained.
\item[(ii)] If only one failure occurs in the first attempt
then after forward or backward propagation of the errors,
either the correct syndrome has been learned or
the extracted parities are inconsistent.
\item[(iii)] If the extracted parities in the first attempt are
inconsistent, then
(i) or (ii)  holds for the second pair.
\end{itemize}
This allows us to label the recovery as having failed if at least one
failure occurred in the network for the first attempt and at least two
failures occurred in the complete recovery network.  (The second
(conditional) attempts' failures have no effect if they are not
actually executed.)
For the purposes of getting a tighter bound on the threshold,
we use the following definitions of failure which can be verified
to be necessary for the recovery network to indentify an incorrect
syndrome or to not indentify a syndrome at all.

\noindent{\bf Definition:}
A syndrome bit extraction has failed if it contains a failure in a
location other than the second cat state preparation attempt.  A
recovery network has failed if it contains at least two failures of
syndrome bit extractions satisfying: The two failures are in different
halves of the first recovery attempt, or they are in
different recovery attempts.

Errors associated with a successful recovery are now propagated
to the previous or the following encoded gate as needed to
have the successful recovery look like an error-free recovery
attempt.

Any encoded gate for which one of the associated recovery networks
has failed, is considered a failed gate.
After having determined which gates failed because of a failed
recovery, we must determine which gates failed because of having
introduced more than one error {\em after} the recovery networks. These are
errors which may have arisen during the step which restores the
syndrome to $0$, or during the encoded operation, or have been
propagated from a successful recovery
or from a subsequent successful gate.

\noindent{\bf Definition:}
An encoded gate has failed if one of the
associated recoveries failed, or they succeeded and there are at least
two failures among the syndrome bit extractions of the second syndrome
measurement of one of its first recovery attempts,
among the operations involved in
restoring the syndrome to $0$, in the encoded operation itself, and in the
first recovery attempt of a following successful
gate. 

Since determining whether a gate failed may require knowing the
following gate's status, this definition requires determining failure
backwards in time so that when a given gate is considered, the
following gate's status is indeed known. Simpler definitions
with the property that failure is determined by considering
errors only in a small neighborhood of an encoded gate
are possible but lead to somewhat worse threshold estimates.

There are two properties of the failure declaration that will be
required for the analysis.  The first guarantees that if after
propagation of errors there is an induced error in the encoded gate,
the gate is determined to have failed. The second ensures that the
failed gates can be associated with disjoint pairs of error locations
which, according to the definition, contributed to the declaration of
failure.

\subsection{Thresholds for the normalizer group}

We begin with the analysis for the stochastic error model and then
explain how it generalizes to the monotonic non-stochastic model.
The analysis exploits the subadditivity property of probabilities over unions
of events to estimate the probability that a given $k$ gates at the
next level are considered failed. To do so requires calculating the
number of possible minimal sets of error locations that can cause a
given gate to fail. In our case, these sets are pairs of
locations. Suppose that the total number of such pairs for a gate is
$f$. By assumption, the probability for a given such pair to have
failed is bounded by $p^2$, where $p$ is the failure probability at the
current level. Thus the probability that the next level gate under
consideration is declared a failure is bounded by $f p^2$. Because of
the second property of the failure definition, the failure of $k$ next
level gates requires $2k$ failures at this level, with $k$ disjoint
pairs of these failures contributing to the failure of each of the $k$
gates. Thus the probability of failure of these gates is bounded by
$(f p^2)^k$, so the new failure probability for the stochastic error model
is bounded by $f p^2$.

It remains to determine the number $f$ of minimal pairs which can
cause a gate to fail. The actual value of $f$ depends to some extent
on the gate. We will obtain an upper bound, and consider memory error
locations separately from other locations.  Excluding the error
locations in the second cat state preparation attempts, the number of
error locations in the network for extracting a bit of the syndrome is
$(\ooo,\mmm)$ operational and memory error locations, respectively.
In the case where memory errors are significant, we assume that
all operations are efficiently pipelined and that conditionally
executed recovery attempts do not cause excessive delays
in other operations. Thus we get a total of
$(6*\ooo+7,6*\mmm) = (\xxx,\yyy)$ locations in a single syndrome extraction
network (there are $6$ syndrome bits to compute and one encoded
Hadamard transform is used in each extraction attempt to switch
between the $\ket{0}/\ket{1}$ and the $\ket{+}/\ket{-}$ basis).  The
number of pairs which can lead to failure of a recovery network is
therefore given by $\xxx^2 + (2*\xxx)^2 = \zzz$ and $\xpyyy+ (2*\xpyyy)^2 =
\mzzz$ with and without memory locations taken into account,
respectively (obtained by counting pairs
of locations with one in the first
and the other in the second syndrome extraction attempt,
and those with the one in the first
and the other in the second recovery attempt).
For state preparation, this increases to $\uuu$ and
$\vvv$, respectively (because of the additional syndrome
bit required for each syndrome extraction).
For two qubit operations there are two
recovery networks, which gives $2*\zzz$ and $2*\mzzz$ minimal failure
sets.  The number of pairs of error locations that can contribute
toward gate failure if the recovery networks succeed can be computed
similarly.  There are up to $14+7=21$ error locations associated with the
encoded operation and the step which resets the syndrome to $0$. (The
maximum occurs for the controlled-not, with $7$ operations
in the encoded controlled-not and up to $14$ required to cancel
errors determined by the recovery network). By
removing error pairs in the following recoveries which would cause
those to fail, we get ${6*\xxx+21\choose 2} - 6*\xxx^2 = \ppp$ and
${6*\xpyyy+21\choose 2} - 6*\xpyyy^2 = \qqq$ for two-qubit gates.
(The pair has to come from the second syndrome extraction attempt
of the gate's recovery networks, from the error-correction,
the encoded operation or the first recovery attempt of a subsequent
gate. Those pairs which arise from the subsequent recoveries
and cause those to fail have been subtracted, as well as those
which introduce only one error in each qubyte.). Other gates have
fewer such error pairs. By
adding the values for the two qubit gates, we get a bound of $f\leq \offf$
and $f\leq \mfff$, respectively. This gives threshold error bounds
of about $\ottt\;10^{-6}$ and $\mttt\;10^{-6}$, respectively.

The monotonic error model requires that for some $C$ and error
parameter $p$, any summand of those error operators contributing to
the failure of a given $k$ gates has strength at most $Cp^k$. Because
of this assumption, the stochastic analysis above can be used almost
verbatim for this model. This is based on the observation that all the
estimates given there are based on bounding the probability of a set
of error events by representing it as a union of events, each of which
fits the requirements for satisfying a bound of the form $Cp^k$.
Thus, the strength of the sum of events which contributes to the
failure of a given $k$ next level gates is bounded by $C (f
p^2)^k$. It can be seen that the requisite monotonicity condition is
also preserved at the next level. Consequently, the bound for the
threshold has the same value, but for strength rather than
probabilities. The quantity $C$ only affects the overhead required to
achieve a desired accuracy of the computation at the uppermost level.

\subsection{Extension to a complete set of operations}

To obtain a complete set of operations it suffices adding the encoded
$\ket{\pi/8}$ preparation gate where needed at the uppermost
(computational) level. To obtain the threshold requirements for this
gate we bound its error given the error behavior of the $\ket{\pi/8}$
gates at the previous level.  The $\ket{\pi/8}$ preparation step fails
if one of its two recovery networks fail, or if there are at least two
errors introduced into the encoded state by propagation, potentially
from a subsequent successful gate. Conservatively, failure not
attributable to the two recovery networks can be reduced to errors in
the first attempts at generating the two cat states for the controlled
encoded Hadamard transforms, the implementation of the controlled
Hadamard transforms (including it's two required $\ket{\pi/8}$
states), the second syndrome extraction in the first attempts at
recovery in each recovery network, or the first attempt at recovery in
the subsequent gate. The computations are similar
to the ones given previously. The number of pairs of errors in
this set not leading to failure of a recovery network is
$\orttt$ and $\mrttt$, with memory errors
taken into account. The number of pairs of errors leading to failure of
one of the two recovery networks is $\oettt$ and $\mettt$, respectively.
Adding these leads to slightly worse thresholds of $\pittt\;10^{-6}$
and $\mpittt\;10^{-6}$, respectively.

\subsection{Leakage errors}

The fault tolerant networks analyzed above are not guaranteed to be
able to suppress leakage errors. To do that one can use a stop leak
gate for qubits which (in the absence of noise) has the property that
a state in the computational space is untouched, while any amplitude
outside this space is irreversibly returned to the computation.  Such
stop leak gates must be inserted before each attempt at extracting the
syndrome bits at the physical level, thus introducing a number of
additional error locations that must be accounted for.  Using these
gates at the physical level changes the interpretation of an encoded
qubit and the behavior of the encoded gates.  The abstract particle
definition of the encoded qubit is modified so that single
leakage events do not destroy the encoded information.  The encoded
gates act correctly on states where at most one supporting qubit has
lost its amplitude. If more qubits lose amplitude, that constitutes a
leakage event at the next level.  However, stop leak measures are no
longer explicitly needed, since each gate's action has been extended
so that (in the absence of an excess of internal errors after error
propagation) it also returns the amplitude to the encoded
computational space. Note that although error propagation cannot be
handled quite as algebraically as is possible with the standard error
group, it is still possible to move leakage errors to
other locations, but with the new failure operator
no longer explicitly determined.  Alternatively, one can linearly
represent all possible operators using standard errors
for higher dimensional spaces. One such approach involves artificially
splitting an extension of the full Hilbert space available to a
physical qubit into $Q\otimes L$ such that $Q\otimes \ket{0}$ is the
computationally useful space. This is akin to the abstract particle
representation of the encoded information, with $L$ playing the role
of the syndrome space. The stop leak gate behaves like a $0$-error
correcting recovery operator.

\subsection{Overheads}

Since we have made no attempt at optimizing the overheads involved
with coding and concatenation we only provide asymptotic estimates for
the method described.  The implementation of each encoded gate
requires a constant amount of resources at the previous level.  Let
$K$ be a bound on the total number of qubits and steps at the previous
level per encoded gate. The physical resources required for a
computational gate is bounded by $K^h$, where $h$ is the number of
levels in the concatenation.  If the number of computational gates
required is $n$, and the desired probability of failure of the
computation is $q$, then the number of levels is sufficient provided
that $n (fp)^{2^h} < q$, where $p$ is the physical probability of
failure.  This gives $h > \log_2(\log_{1/(fp)}(n/q))$ and an overhead
per computational operation of at least $K^h >
(\log_{2}(n/q)/\log_2(1/(fp)))^{\log_2(K)}$, which is polylogarithmic
in $n$ and $1/q$. With the assumption that measured qubits can be
freely reused this is also a bound on the number of qubits required to
support each computational qubit. Otherwise, this overhead can
increase by an additional factor related to the the number of gates
the qubit is involved in.  In our case $\log_2(K)$ is about $10$,
which can be quite daunting in practice.  Substantial asymptotic
improvements can be obtained by using different codes at higher levels
and by encoding multiple qubits using single blocks.

\section{Conclusion}
\label{section:conclusion}

We have demonstrated that quantum computation with classical input can
be performed arbitrarily accurately provided that the noise per
operation is sufficiently small and satisfies suitable independence
assumptions. The implementation of fault tolerant quantum computation
is straightforward. Since the overheads are asymptotically well
behaved, the threshold results demonstrate that quantum computation is
possible in the presence of physically reasonable sources of noise.

Threshold results have been obtained independently by
Kitaev~\cite{kitaev:qc1996a} and Aharonov and
Ben-Or~\cite{aharonov:qc1996a}. Both Kitaev and Aharonov and Ben-Or
analyze independent stochastic error models and obtain
completeness of operations by adopting Shor's implementation of the
Toffoli gate. Kitaev uses a different method for fault tolerantly
extracting the syndrome.  His method is less efficient and
consequently yields substantially worse thresholds. Aharonov and
Ben-Or provide an analysis which does not require accurate classical
computation for syndrome calculations and estimate a threshold of
around $10^{-6}$ for the independent stochastic error model. There may
be some cases where this extension is needed, for example when
performing stochastically parallel quantum computations such as those
envisioned for NMR quantum
computation~\cite{cory:qc1996a,chuang:qc1997a}.  Our
techniques for generalizing arguments from stochastic error models to
coherent error models can in principle be used to extend their
analysis.

Not only do the threshold theorems show that quantum
computation is possible in principle, but they demonstrate
that the apparent distance limitations of quantum cryptography
can be overcome. It suffices to be able to transmit the
state of qubits over some reasonable distance before
recovery operations must be applied to avoid loss of encoded
information. This does of course require compatible physical
realizations of qubits for transmission over channels
and for manipulation in a quantum computer.

The actual values of the thresholds we have obtained are rigorous, but
overly pessimistic in several ways. First, the error models used are
the most adversarial still satisfying independence assumptions. We
assume that the error behavior at each location is the worst possible
for the network subject only to bounds on the strength of
correlations. In practice, we need not worry about such adversarial
error behavior and the actual error types at the physical level are
likely to be much more constrained than assumed by our error-blind
analysis. That known error behavior can be exploited to 
reduce error has been demonstrated in a specific example by Pellizzari,
Cirac, Pellizzari and Zoller~\cite{cirac:qc1996a}. In addition,
simulations due to Preskill's team (private communication) and
Zalka~\cite{zalka:qc1996a} suggest that for the depolarizing channel,
thresholds are substantially better than suggested by our
calculations. Second, we have made no attempt to optimize the
implementation of fault tolerance. Suggestions for optimization can be
found in~\cite{knill:qc1996j,zalka:qc1996a}.  Finally, it is clear
that our failure definitions are excessively conservative, and a
substantial fraction of error pairs which lead us to consider a gate
failed in fact do not induce an error at the next level.
Nevertheless, the results suggest that over-rotation and other
errors not representable stochastically must be controlled more
carefully than stochastic noise.

Whether fault tolerant quantum computation can be implemented in
practice remains to be seen. However, the results obtained here
show that in principle noise of a level below the error threshold 
is not an obstacle for quantum computation.


\bibliographystyle{prsty}

\vfill\pagebreak\onecolumn

\mbox{}
\vfill

\begin{table}
\centering
\tabcolsep .0in
\begin{tabular}{|@{\hspace{.1in}}c@{\hspace{.1in}}|@{\hspace{.3in}}c@{\hspace{.3in}}|@{\hspace{.2in}}c@{\hspace{.2in}}|@{\hspace{.3in}}c@{\hspace{.3in}}|}
\hline
Names & Symbols & Action & Gate icon\\
\hline
\hline
not, bit flip & $-i\sigma_y, N$ &
$\left(\begin{array}{cc}
0&1\\1&0
\end{array}\right)$ & \raisebox{-.1in}{\psfig{figure=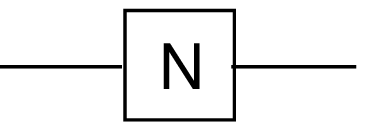,width=.8in,height=.25in}}
\\
\hline
sign flip & $\sigma_z, S$ &
$\left(\begin{array}{cc}
1&0\\0&-1
\end{array}\right)$&\raisebox{-.1in}{\psfig{figure=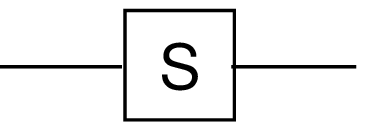,width=.8in,height=.25in}}
\\
\hline
$i$-phase shift & $S_i$ &
$\left(\begin{array}{cc}
1&0\\0&i
\end{array}\right)$&\raisebox{-.1in}{\psfig{figure=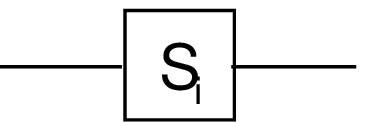,width=.8in,height=.25in}}
\\
\hline
Hadamard rotation & $H$ &
$\left(\begin{array}{cc}
1&1\\1&-1
\end{array}\right)$&\raisebox{-.1in}{\psfig{figure=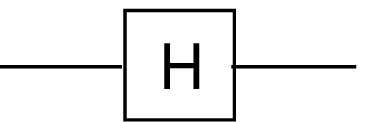,width=.8in,height=.25in}}
\\
\hline
controlled-not, xor & $N_2$ &
$\ket{x}\ket{y} \rightarrow \ket{x}\ket{x+y\bmod{2}}$ &\raisebox{-.2in}[.4in][.3in]{\psfig{figure=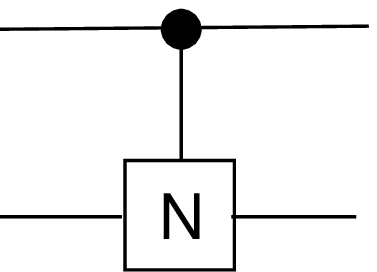,width=.8in,height=.5in}}
\\
\hline
Toffoli gate & $T$ &
$\ket{x}\ket{y}\ket{z} \rightarrow \ket{x}\ket{y}\ket{z+xy\bmod{2}}$ &
\raisebox{-.2in}[.7in][.3in]{\psfig{figure=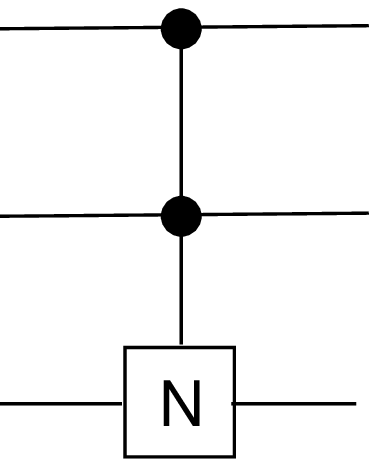,width=.8in,height=.8in}}
\\
\hline
Preparation of $\ket{\psi}$ & & &
\raisebox{-.1in}[.3in][.2in]{\psfig{figure=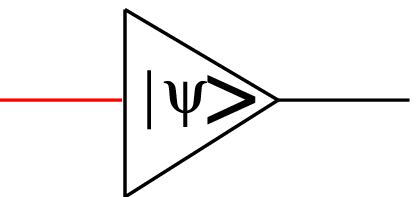,width=.8in,height=.3in}}
\\
\hline
Measurement of a qubit & & &\raisebox{-.1in}[.3in][.2in]{\psfig{figure=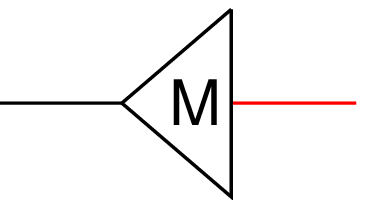,width=.8in,height=.3in}}
\\
\hline
\end{tabular}
\caption{Elementary unitary operations on one to three qubits.}
\label{table:ops}
\end{table}

\begin{table}
\begin{center}
\begin{eqnarray*}
SN &=& -NS\\
S H &=& H N\\
N H &=& H S\\
S S_i &=& S_i S\\
N S_i &=& -i S_i N\\
(I\otimes S) N_2 &=& N_2 (S\otimes S)\\
(S\otimes I) N_2 &=& - N_2 (S\otimes I)\\
(I\otimes N) N_2 &=& N_2 (I\otimes N)\\
(N\otimes I) N_2 &=& N_2 (N\otimes N)
\end{eqnarray*}
\end{center}
\caption{Error propagation identities obtained by conjugation ($EV = V (V^\dagger E V)$).}
\label{table:conjugation-errors}
\end{table}

\vfill

\pagebreak
\mbox{}

\vfill

\vfill

\pagebreak
\mbox{}

\vfill 
 
\vfill

\pagebreak

\mbox{}
\vfill
 
\vfill

\pagebreak

\mbox{}
\vfill

\vfill

\pagebreak
\mbox{}

\vfill

\vfill
\pagebreak
\mbox{}
\vfill

\vfill

\pagebreak
\mbox{}

\vfill

\vfill

\pagebreak

\vfill\eject

\end{document}